\documentclass[12pt]{article}

\usepackage{amsmath,amsthm,upref}
\usepackage{amssymb,amsbsy}
\usepackage{amsfonts}
\usepackage{cite}

\def\dint{\mathop{\displaystyle \int}}

\author{{\bf Steven Duplij and
Albert Kotvytskiy}\\[5pt]
{\it V.N. Karazin
Kharkov National University}\\ {\it Svoboda Sq. 4, Kharkov 61077, Ukraine}\\
{\small steven.a.duplij@univer.kharkov.ua, albert.t.kotwicki@univer.kharkov.ua}\\
{\small http://webusers.physics.umn.edu/\~{}duplij}}

\title{\textbf{Coincidence limit and generalized interaction
term structure\\ in multigravity}}

\date{2 December,  2007}

\begin{document}
\maketitle

\begin{abstract}

Generalized structure of the interaction term of multigravity
is analyzed in detail. The coincidence limit of any multigravity theory is
defined and the compatibility equation for the interaction potential is
derived which is studied in the weak perturbation limit of metric. The most
general properties of the invariant volume and the scalar potential of
multigravity are investigated. The general formula for multigravity
invariant volume using three means (arithmetic, geometric and harmonic) is
derived. The Pauli-Fierz mass term for bigravity in the weak field limit is
obtained. 

\end{abstract}

\newpage

\section{Introduction}

The multigravity extension of General Relativity (in first papers it was
called \textquotedblleft \textit{f-g} theory\textquotedblright\ or
\textquotedblleft strong gravity\textquotedblright\ \cite%
{ish/sal/str,aic/man/urb,aic}) is important both from theoretical
constructions (quantum gravity and branes \cite%
{kog/ros,kog/mou/pap/ros,kog/mou/pap}, discrete dimensions \cite%
{def/mou1,def/mou2}, renormalization \cite{gar1}, massive gravity \cite{bla}
etc.) and experimental facts (dark matter and dark energy \cite%
{han,gri/pav,dub/tin/tka}, cosmic acceleration \cite{dam/kog/pap,def/dva/gab}
etc.). In this respect it is worthwhile to consider non-linear formulation
of multigravity \cite{dam/kog}. The shape of interaction term plays the most
crucial role in constructing models.

The goal of this paper is to consider the generalized structure of the
interaction term in detail (see also \cite{dup/kotv2004}). That is, we
introduce the coincidence limit of a multigravity theory and obtain the
compatibility equation for the interaction potential and analyze it in the
weak perturbation limit. Note that a particular case of our general
construction, a \textquotedblleft perturbative limit\textquotedblright\
which corresponds to critical points of interacting potential and depends
from their special form of interaction potential, was considered in \cite%
{dam/kog} for bigravity only. Here we propose the multigravity
generalizations and do not consider any restrictions on the metric, as in
\cite{dam/kog} (where only spaces with constant curvature were considered).

Also we study the most general properties of invariant volume in the
interaction term and the scalar potential of multigravity. We generalize the
invariant volume for multigravity for three means and obtain the Pauli-Fierz
mass term \cite{fie/pau} for bigravity in the weak field limit \cite{dam/kog}%
, as an example.

\section{Multigravity and the coincidence limit}

We consider several Universes (labelled by $i=1,\ldots N$) each described by
the metric $\mathsf{g}_{\mu \nu }^{\left( i\right) }$ (we use the signature $%
+---$), the set of matter fields $\Phi ^{\left( i\right) }$ (scalar,
spinorial, vector ones) and the action%
\begin{equation}
S_{G\left( i\right) }=\int d\Omega ^{\left( i\right) }\left[ F^{\left(
i\right) }(\mathsf{g}^{\left( i\right) })+L\left( \mathsf{g}^{\left(
i\right) },\Phi ^{\left( i\right) }\right) \right] ,  \label{s}
\end{equation}%
where $d\Omega ^{\left( i\right) }=d^{4}x\sqrt{g^{\left( i\right) }}$, $%
g^{\left( i\right) }=-\det \left( \mathsf{g}_{\mu \nu }^{\left( i\right)
}\right) >0$ (distinguishing $g^{\left( i\right) }$ as a positive number and
$\mathsf{g}^{\left( i\right) }$ as a tensor) is the invariant volume and $%
F^{\left( i\right) }(\mathsf{g}^{\left( i\right) })$ is pure gravity
Lagrangian of $i$-Universe, $L\left( \mathsf{g}^{\left( i\right) },\Phi
^{\left( i\right) }\right) $ describes coupling of matter fields and
gravity. In the concept of Weakly Coupled Worlds \cite{dam/kog} due to the
no-go theorem of \cite{bou/dam/gua/hen} the only consistent nonlinear theory
of $N$ massless gravitons is the sum of decoupled gravity actions (\ref{s})%
\begin{equation}
S_{0}=\sum_{i}^{N}S_{G\left( i\right) }
\end{equation}%
which has the huge symmetry $\prod_{i}^{N}\mathrm{diff}_{\left( i\right) }$
(each $\mathrm{diff}_{\left( i\right) }$ acts on its metric $\mathsf{g}_{\mu
\nu }^{\left( i\right) }$ and matter fields $\Phi ^{\left( i\right) }$). The
full action of multigravity, as Weakly Coupled Worlds mixing by their
gravitational fields only, is

\begin{equation}
S_{mG}=\sum_{i}^{N}\int d\Omega ^{\left( i\right) }\left[ F^{\left( i\right)
}(\mathsf{g}^{\left( i\right) })+L\left( \mathsf{g}^{\left( i\right) },\Phi
^{\left( i\right) }\right) \right] +\int d^{4}xW(\mathsf{g}^{\left( 1\right)
},\mathsf{g}^{\left( 2\right) },\ldots \mathsf{g}^{\left( N\right) }),
\label{sm}
\end{equation}%
where $W(\mathsf{g}^{\left( 1\right) },\mathsf{g}^{\left( 2\right) },\ldots
\mathsf{g}^{\left( N\right) })$ is the interaction term which is a scalar
density made up from metrics taken at the same point, i.e. in ultralocal
limit \cite{dam/kog}. The symmetry of (\ref{sm}) reduces to only one
diffeomorphism, because of the no-go theorem \cite{bou/dam/gua/hen}.
Therefore, it is interesting to consider the case when also the Universities
are described by the same metric. So let us introduce the coincidence limit,
when $\mathsf{g}_{\mu \nu }^{\left( 1\right) }=\mathsf{g}_{\mu \nu }^{\left(
2\right) }=\ldots =\mathsf{g}_{\mu \nu }^{\left( N\right) }\equiv \mathsf{g}%
_{\mu \nu }$. In case of the absence of interaction ($W=0$) and matter, we
have%
\begin{equation}
S_{0}=\int d\Omega \sum_{i}^{N}F^{\left( i\right) }(\mathsf{g}).
\end{equation}%
If $F^{\left( 1\right) }(\mathsf{g})=F^{\left( 2\right) }(\mathsf{g})=\ldots
=F^{\left( N\right) }(\mathsf{g})\equiv F(\mathsf{g})$, then $S_{0}=N\int
d\Omega F(\mathsf{g}_{\mu \nu })$, and therefore noninteracting full theory
coincides with the initial one. But in the case of interacting theory and
moreover nonvanishing interacting term in the coincidence limit the
multigravity can be equivalent to some effective gravity theory described by
the effective metric $\mathsf{\tilde{g}}_{\mu \nu }$ and effective function $%
\tilde{F}\left( \mathsf{\tilde{g}}\right) $. Thus we arrive to the
compatibility equation%
\begin{equation}
\sqrt{g}\left( F(\mathsf{g})+U\left( \mathsf{g}\right) \right) =\sqrt{\tilde{%
g}}F(\mathsf{\tilde{g}}),  \label{comp}
\end{equation}%
where $\sqrt{g}U\left( \mathsf{g}\right) =W(\mathsf{g},\mathsf{g},\ldots
\mathsf{g})\neq 0$, and all functions are taken in the same `point'. The
equation (\ref{comp}) is defined up to covariant divergence of any function,
because it will not contribute to the equations of motion. In \cite%
{ish/sal/str,aic/man/urb,dam/kog} the only case $W(\mathsf{g},\mathsf{g}%
,\ldots \mathsf{g})=0$ ($U\left( \mathsf{g}\right) =0$) was considered, and
the compatibility equation has the trivial solution $\mathsf{\tilde{g}}=%
\mathsf{g}$ only. Here we extend the consideration to nonvanishing $U\left(
\mathsf{g}\right) $, which allows us to obtain possible nontrivial
solutions. The physical sense of the compatibility equation (\ref{comp}) is
treatment of two equal interacting Universes (having the same function $F$)
in the limit of coinciding metric tensors, as some \textquotedblleft
effective\textquotedblright\ Universe described by this function $F$, but
another metric tensor $\mathsf{\tilde{g}}$.

In general case the formal solution of the compatibility equation (\ref{comp}%
) can be presented as%
\begin{equation*}
\mathsf{\tilde{g}}_{\mu \nu }=\Phi _{\mu \nu }\left( \mathsf{g},U\left(
\mathsf{g}\right) \right) ,
\end{equation*}%
where the function $\Phi _{\mu \nu }$ is a symmetric covariant tensor
determining the transformation $\mathsf{g}_{\mu \nu }\rightarrow \mathsf{%
\tilde{g}}_{\mu \nu }$.

Let us solve the compatibility equation in the simplest case: small fields
expansion%
\begin{equation}
\mathsf{\tilde{g}}_{\mu \nu }=\mathsf{g}_{\mu \nu }+\mathsf{p}_{\mu \nu }.
\label{gp}
\end{equation}

We note that here we consider $\mathsf{g}_{\mu \nu }$ as an arbitrary
metric, but not necessarily flat space metric $\mathsf{g}_{\mu \nu }\neq
\eta _{\mu \nu }$. In the first order of $\mathsf{p}_{\mu \nu }$ for
determinants we derive%
\begin{eqnarray}
\det \left( \mathsf{\tilde{g}}\right)  &=&\det \left( \mathsf{g}\right) +%
\mathsf{p}_{\alpha \beta }\mathsf{K}^{\alpha \beta }\left( \mathsf{g}\right)
,  \notag \\
\mathsf{K}^{\alpha \beta }\left( \mathsf{g}\right)  &=&\varepsilon ^{\mu \nu
\rho \sigma }\left( \delta _{0}^{\alpha }\delta _{\mu }^{\beta }\mathsf{g}%
_{1\nu }\mathsf{g}_{2\rho }\mathsf{g}_{3\sigma }+\delta _{1}^{\alpha }\delta
_{\nu }^{\beta }\mathsf{g}_{0\mu }\mathsf{g}_{2\rho }\mathsf{g}_{3\sigma
}\right.   \notag \\
&&\left. +\delta _{2}^{\alpha }\delta _{\rho }^{\beta }\mathsf{g}_{0\mu }%
\mathsf{g}_{1\nu }\mathsf{g}_{3\sigma }+\delta _{3}^{\alpha }\delta _{\sigma
}^{\beta }\mathsf{g}_{0\mu }\mathsf{g}_{1\nu }\mathsf{g}_{2\rho }\right) .
\label{k}
\end{eqnarray}

If we consider expansion around Minkowski metric $\mathsf{g}_{\mu \nu }=\eta
_{\mu \nu }$, then $\mathsf{K}^{\alpha \beta }\left( \mathsf{g}\right)
=-\eta ^{\alpha \beta }$ and $\tilde{g}=-\det \left( \mathsf{\tilde{g}}%
\right) =1+\mathrm{Tr}\ \mathsf{p}$, where $\mathrm{Tr}\ \mathsf{p}\equiv
\mathsf{p}_{\alpha \beta }\eta ^{\alpha \beta }$. In general case, after
substitution of (\ref{k}) into the main compatibility equation (\ref{comp}),
we obtain%
\begin{eqnarray}
U\left( \mathsf{g}\right)  &=&\left( \dfrac{\partial F\left( \mathsf{g}%
\right) }{\partial \mathsf{g}_{\mu \nu }}-\dfrac{1}{2\sqrt{g}}F\left(
\mathsf{g}\right) \mathsf{K}^{\mu \nu }\left( \mathsf{g}\right) \right)
\mathsf{p}_{\mu \nu }  \notag \\
&&+\dfrac{\partial F\left( \mathsf{g}\right) }{\partial \mathsf{g}_{\mu \nu
,\rho }}\mathsf{p}_{\mu \nu ,\rho }+\dfrac{\partial F\left( \mathsf{g}%
\right) }{\partial \mathsf{g}_{\mu \nu ,\rho \sigma }}\mathsf{p}_{\mu \nu
,\rho \sigma }+\ldots ,  \label{u}
\end{eqnarray}%
where \textquotedblleft \ldots \textquotedblright\ denote similar
derivatives by higher than two derivatives of $\mathsf{g}_{\mu \nu }$ terms.

So any multigravity model (\ref{s}) induces the interaction term which in
the coincidence limit has the form (\ref{u}). On the other hand, the
relation (\ref{u}) can be considered as an equation for $\mathsf{p}_{\mu \nu
}$, and therefore we can determine an effective metric $\mathsf{\tilde{g}}%
_{\mu \nu }$ of gravity theory, which is equivalent to a given multigravity
in the coincidence limit, for any interaction term.

In most cases $F\left( \mathsf{g}\right) $ is a function of Riemann
curvature $R_{\mu \nu \rho \sigma }\left( \mathsf{g}\right) $ which contains
only up to 2 derivatives of the metric, and so the higher terms in (\ref{u})
denoted by ``$\ldots $'' will not appear. A general polynomial shape of such $%
F\left( \mathsf{g}\right) $ is%
\begin{equation*}
F\left( \mathsf{g}\right) =\hat{F}\left( R_{\mu \nu \rho \sigma }\left(
\mathsf{g}\right) \right) =A\cdot R^{n}\left( \mathsf{g}\right) +B\cdot
R_{\mu \nu }^{m}\left( \mathsf{g}\right) +C\cdot R_{\mu \nu \rho \sigma
}^{r}\left( \mathsf{g}\right) ,
\end{equation*}%
where $A,B,C$ are constants and%
\begin{eqnarray*}
R_{\ \nu \rho \sigma }^{\mu }\left( \mathsf{g}\right)  &=&\Gamma _{\ \nu
\sigma ,\rho }^{\mu }\left( \mathsf{g}\right) -\Gamma _{\ \nu \rho ,\sigma
}^{\mu }\left( \mathsf{g}\right) +\Gamma _{\ \tau \rho }^{\mu }\left(
\mathsf{g}\right) \Gamma _{\ \nu \sigma }^{\tau }\left( \mathsf{g}\right)
-\Gamma _{\ \tau \sigma }^{\mu }\left( \mathsf{g}\right) \Gamma _{\ \nu \rho
}^{\tau }\left( \mathsf{g}\right) , \\
\Gamma _{\ \nu \rho }^{\mu }\left( \mathsf{g}\right)  &=&\dfrac{1}{2}\mathsf{%
g}^{\mu \sigma }\left( \mathsf{g}_{\sigma \nu ,\rho }+\mathsf{g}_{\sigma
\rho ,\nu }-\mathsf{g}_{\nu \rho ,\sigma }\right) , \\
R_{\mu \nu }\left( \mathsf{g}\right)  &=&R_{\ \mu \rho \nu }^{\rho }\left(
\mathsf{g}\right) ,\ \ \ \ \ \ R\left( \mathsf{g}\right) =\mathsf{g}^{\mu
\nu }R_{\mu \nu }\left( \mathsf{g}\right) .
\end{eqnarray*}

The standard Einstein gravity corresponds to $F\left( \mathsf{g}\right)
=A_{Einsten}\cdot R\left( \mathsf{g}\right) $ \cite{lan/lif2}. In this case
and using (\ref{gp}) we have (note the absence of the first derivatives of $%
\mathsf{g}_{\mu \nu }$)%
\begin{eqnarray}
U_{Einsten}\left( \mathsf{g}\right)  &=&A_{Einsten}\left[ \left( \dfrac{%
\partial R\left( \mathsf{g}\right) }{\partial \mathsf{g}_{\mu \nu }}-\dfrac{1%
}{2\sqrt{g}}R\left( \mathsf{g}\right) \mathsf{K}^{\mu \nu }\left( \mathsf{g}%
\right) \right) \mathsf{p}_{\mu \nu }\right.   \notag \\
&&\left. +\dfrac{\partial R\left( \mathsf{g}\right) }{\partial \mathsf{g}%
_{\mu \nu ,\rho }}\mathsf{p}_{\mu \nu ,\rho }+\dfrac{\partial R\left(
\mathsf{g}\right) }{\partial \mathsf{g}_{\mu \nu ,\rho \sigma }}\mathsf{p}%
_{\mu \nu ,\rho \sigma }\right] .
\end{eqnarray}

It is convenient to use covariant derivatives by $\mathsf{g}_{\mu \nu }$,
then%
\begin{eqnarray*}
\widetilde{\Gamma }_{\nu \rho }^{\mu } &=&\Gamma _{\nu \rho }^{\mu }+\frac{1%
}{2}\mathsf{g}^{\mu \sigma }(\mathsf{p}_{\sigma \nu ;\rho }+\mathsf{p}%
_{\sigma \rho ;\nu }-\mathsf{p}_{\nu \rho ;\sigma }), \\
\widetilde{R}_{\nu \rho \sigma }^{\mu } &=&R_{\nu \rho \sigma }^{\mu }+\frac{%
1}{2}\mathsf{g}^{\mu \alpha }(\mathsf{p}_{\alpha \nu ;\sigma \rho }+\mathsf{p%
}_{\alpha \sigma ;\nu \rho }-\mathsf{p}_{\nu \sigma ;\alpha \rho }-\mathsf{p}%
_{\alpha \nu ;\rho \sigma }-\mathsf{p}_{\alpha \rho ;\nu \sigma }+\mathsf{p}%
_{\nu \rho ;\alpha \sigma }), \\
\widetilde{R} &\equiv &\widetilde{g}^{\nu \sigma }\widetilde{R}_{\nu \mu
\sigma }^{\mu }=R-\mathsf{p}^{\alpha \beta }R_{\alpha \beta }-\square
\mathsf{p}_{\alpha ;\beta }^{\alpha \quad \beta } \\
&&+\frac{1}{2}\mathsf{g}^{\alpha \beta }\mathsf{g}^{\mu \nu }(\mathsf{p}%
_{\beta \mu ;\nu \alpha }-\mathsf{p}_{\beta \mu ;\alpha \nu }+\mathsf{p}%
_{\beta \nu ;\mu \alpha }+\mathsf{p}_{\mu \alpha ;\beta \nu }),
\end{eqnarray*}%
where $\square $ is covariant D'Alambertian defined as $\square =\nabla
_{\mu }\nabla ^{\mu }$ and $\nabla _{\mu }$ is covariant derivative by $%
\mathsf{g}_{\mu \nu }$, i.e. $\square \mathsf{p}\equiv \mathsf{p}_{\ \alpha
;\beta }^{\alpha \quad \beta }$.

After substitution to (\ref{comp}) we obtain

\begin{eqnarray}
&&\dint \widetilde{R}\sqrt{\widetilde{g}}d^{4}x=\dint \sqrt{g}d^{4}xR+\dint
\sqrt{g}d^{4}x\left[ -\mathsf{p}^{\alpha \beta }R_{\alpha \beta }-\square
\mathsf{p}\right.  \notag \\
&&\left. +\frac{1}{2}\mathsf{g}^{\alpha \beta }\mathsf{g}^{\mu \nu }(\mathsf{%
p}_{\beta \mu ;\nu \alpha }-\mathsf{p}_{\beta \mu ;\alpha \nu }+\mathsf{p}%
_{\beta \nu ;\mu \alpha }+\mathsf{p}_{\mu \alpha ;\beta \nu })-\frac{R}{2%
\sqrt{g}}\mathsf{p}_{\alpha \beta }\mathsf{F}^{\alpha \beta }\right] .
\end{eqnarray}

\section{Generalized invariant volume in multigravity}

In consideration of the interaction term of multigravity it is important to
choose consistently the invariant volume which in coincidence limit
transforms to the standard invariant volume $d^{4}x\sqrt{g}$. For
simplicity, first we consider the bigravity case \cite{dam/kog}.

Note that $d^{4}xW(\mathsf{g}^{\left( 1\right) },\mathsf{g}^{\left( 2\right)
})$ is a scalar, while $d^{4}x$ and $W(\mathsf{g}^{\left( 1\right) },\mathsf{%
g}^{\left( 2\right) })\ $are the scalar densities of opposite weights. By
analogy with usual invariant volume $d\Omega =d^{4}x\sqrt{g}$, we can
present $d^{4}xW(\mathsf{g}^{\left( 1\right) },\mathsf{g}^{\left( 2\right)
}) $ as a product $d^{4}x\cdot f\left( \sqrt{g_{1}},\sqrt{g_{2}}\right)
\cdot V(\mathsf{g}^{\left( 1\right) },\mathsf{g}^{\left( 2\right) })$, where
$V(\mathsf{g}^{\left( 1\right) },\mathsf{g}^{\left( 2\right) })$ is a scalar
interaction potential.

Now we demand that the ``interaction'' invariant volume defined by $d\Omega
_{int}=d^{4}xf\left( \sqrt{g_{1}},\sqrt{g_{2}}\right) $ should be a scalar
which in the coincidence limit $\mathsf{g}_{\mu \nu }^{\left( 1\right) }=%
\mathsf{g}_{\mu \nu }^{\left( 2\right) }\equiv \mathsf{g}_{\mu \nu }$ gives
the standard invariant volume $d\Omega _{int}\rightarrow d\Omega $. To
satisfy these conditions we require the following general properties of the
function $f\left( u,v\right) $:

1) Idempotence $f\left( u,u\right) =u$; 2) Monotony; 3) Homogeneity $f\left(
tu,tv\right) =tf\left( u,v\right) $; 4) Symmetry $f\left( u,v\right)
=f\left( v,u\right) $.

From homogeneity and symmetry it follows that $f\left( u,v\right) $ can be
expressed through the function of one variable, the ratio $\dfrac{u}{v}$, as%
\begin{equation}
f\left( u,v\right) =u\cdot f\left( \dfrac{v}{u},1\right) =v\cdot f\left(
\dfrac{u}{v},1\right) =\sqrt{uv}\cdot f\left( \sqrt{\dfrac{u}{v}},\sqrt{%
\dfrac{v}{u}}\right) .  \label{f}
\end{equation}%
Thus, the interaction invariant volume can be presented as%
\begin{equation}
d\Omega _{int}=d^{4}xf\left( \sqrt{g_{1}},\sqrt{g_{2}}\right) =d^{4}x\cdot
\sqrt[4]{g_{1}g_{2}}\cdot f\left( \sqrt[4]{\dfrac{g_{1}}{g_{2}}},\sqrt[4]{%
\dfrac{g_{2}}{g_{1}}}\right) =d^{4}x\cdot \sqrt[4]{g_{1}g_{2}}\cdot \hat{f}%
\left( \dfrac{g_{2}}{g_{1}}\right) .  \label{ff}
\end{equation}%
From symmetry of $f\left( u,v\right) $ it follows that $\hat{f}\left(
u\right) =\hat{f}\left( u^{-1}\right) $.

Let us consider an example. The simplest functions satisfying (\ref{f}) are
usual averages: arithmetic mean, harmonic mean and geometric mean\footnote{%
Usually one considers the geometric mean only (e.g. see \cite{dam/kog}).}.
It is reasonable to consider their linear combination, which gives for the
generalized ``interaction'' invariant volume the following expression%
\begin{eqnarray}
&&d\Omega _{int}\left( a,b,c\right)  =d^{4}xf\left( \sqrt{g_{1}},\sqrt{g_{2}}%
\right)   \notag \\
&&=\frac{d^{4}x}{a+b+c}\left( a\frac{\sqrt{g_{1}}+\sqrt{g_{2}}}{2}+b\sqrt[4]{%
g_{1}g_{2}}+c\frac{2}{\dfrac{1}{\sqrt{g_{1}}}+\dfrac{1}{\sqrt{g_{2}}}}%
\right)   \notag \\
&&=d^{4}x\cdot \sqrt[4]{g_{1}g_{2}}\cdot \frac{1}{a+b+c}\left[ \frac{a}{2}%
\left( \sqrt[4]{\frac{g_{1}}{g_{2}}}+\sqrt[4]{\frac{g_{2}}{g_{1}}}\right)
+b+2c\frac{1}{\sqrt[4]{\frac{g_{1}}{g_{2}}}+\sqrt[4]{\frac{g_{2}}{g_{1}}}}%
\right]   \notag \\
&&=d^{4}x\cdot \sqrt[4]{g_{1}g_{2}}\cdot \frac{1}{a+b+c}\left[ \frac{a}{2}%
\left( y+\dfrac{1}{y}\right) +b+2c\frac{1}{y+\dfrac{1}{y}}\right] ,
\label{n2}
\end{eqnarray}%
where $a$, $b$, $c$ are arbitrary real constants and $y=\sqrt[4]{\frac{g_{1}%
}{g_{2}}}$. Similar formulas are valid for $N$-multigravity%
\begin{eqnarray*}
&&d\Omega _{int}=d^{4}x\cdot f\left( \sqrt{g_{1}},...,\sqrt{g_{N}}\right)  \\
&=&d^{4}x\sqrt[2N]{g_{1}...g_{N}}\cdot f\left( \sqrt[2N]{\frac{g_{1}^{N-1}}{%
g_{2}g_{3}...g_{N}}},\sqrt[2N]{\frac{g_{2}^{N-1}}{g_{1}g_{3}...g_{N}}}...,%
\sqrt[2N]{\frac{g_{N}^{N-1}}{g_{1}g_{2}...g_{N-1}}}\right) .
\end{eqnarray*}%
Evidently, this formula for $N=2$ (bigravity) gives (\ref{ff}).

Let us denote the $N$ arguments of the function $f$ as%
\begin{eqnarray}
y_{1}^{\left( N\right) } &=&\sqrt[2N]{%
g_{1}^{N-1}g_{2}^{-1}g_{3}^{-1}...g_{N}^{-1}},
\notag \\y_{2}^{\left( N\right) }
&=&\sqrt[2N]{g_{1}^{-1}g_{2}^{N-1}g_{3}^{-1}...g_{N}^{-1}},\ldots \notag\\
y_{N}^{\left( N\right) }&=&\sqrt[2N]{%
g_{1}^{-1}g_{2}^{-1}...g_{N-1}^{-1}g_{N}^{N-1}},  \label{yy}
\end{eqnarray}%
which obviously satisfy%
\begin{equation}
y_{1}^{\left( N\right) }\cdot y_{2}^{\left( N\right) }\cdot \ldots \cdot
y_{N}^{\left( N\right) }=1.  \label{y}
\end{equation}%
Therefore the function $f$ has actually $N-1$ independent arguments, and so%
\begin{equation}
d\Omega _{int}=d^{4}x\cdot f\left( \sqrt{g_{1}},...,\sqrt{g_{N}}\right)
=d^{4}x\sqrt[2N]{g_{1}...g_{N}}\cdot \hat{f}\left( y_{1}^{\left( N\right)
},y_{2}^{\left( N\right) },\ldots y_{N}^{\left( N-1\right) }\right) ,
\end{equation}%
which for $N=2$ gives (\ref{ff}). Using the means as in (\ref{n2}) we obtain
its $N$-analog%
\begin{equation}
d\Omega _{int}=d^{4}x\cdot \sqrt[2N]{g_{1}...g_{N}}\cdot \frac{1}{a+b+c}%
\left[ \frac{a}{N}\sum_{i=1}^{N}y_{i}^{\left( N\right) }+b+c\frac{N}{%
\sum_{i=1}^{N}\dfrac{1}{y_{i}^{\left( N\right) }}}\right] ,
\end{equation}%
which can be considered as the most general ``interaction'' invariant volume
for multigravity.

\section{Generalized interaction potential in multigravity}

Let us construct the most general expression for the interaction term in (%
\ref{sm})%
\begin{equation}
S_{int}=\int d^{4}xW(\mathsf{g}^{\left( 1\right) },\mathsf{g}^{\left(
2\right) },\ldots \mathsf{g}^{\left( N\right) }).
\end{equation}%
It is convenient to extract the generalized invariant volume (presented in
previous section)%
\begin{equation}
S_{int}=\int d\Omega _{int}V(\mathsf{g}^{\left( 1\right) },\mathsf{g}%
^{\left( 2\right) },\ldots \mathsf{g}^{\left( N\right) }),
\end{equation}%
where $V(\mathsf{g}^{\left( 1\right) },\mathsf{g}^{\left( 2\right) },\ldots
\mathsf{g}^{\left( N\right) })$ is the scalar interaction potential of
multigravity.

As we noted before, the symmetry of the full action (\ref{sm}) can be
reduced to only one diffeomorphism group which is the diagonal subgroup of
common diffeomorphisms acting on metrics as Lie derivative $\delta \mathsf{g}%
^{\left( i\right) }=\mathcal{L}_{\varepsilon }\mathsf{g}^{\left( i\right) }$
or in manifest form%
\begin{equation}
\delta \mathsf{g}_{\mu \nu }^{\left( i\right) }=\varepsilon ^{\rho }\mathsf{g%
}_{\mu \nu ,\rho }^{\left( i\right) }+\varepsilon _{~,\mu }^{\rho }\mathsf{g}%
_{\rho \nu }^{\left( i\right) }+\varepsilon _{~,\nu }^{\rho }\mathsf{g}_{\mu
\rho }^{\left( i\right) },  \label{dg}
\end{equation}%
where $\varepsilon ^{\rho }$ is the same for all metrics. This symmetry
restricts the shape of the scalar interaction potential: it should depend
from invariant which can be constructed from $N$ metrics $\mathsf{g}_{\mu
\nu }^{\left( i\right) }$.

Let us consider bigravity as an example \cite{dam/kog,dam/kog/pap}. The
scalar potential should depend from invariants of the mixed tensor
\begin{equation}
\mathsf{Y}_{\nu }^{\mu }=\mathsf{g}_{\nu \rho }^{\left( 1\right) }\mathsf{g}%
^{\left( 2\right) \rho \mu },  \label{yg}
\end{equation}%
which can be treated as tensorial analog of the scalar variable $y$ from (%
\ref{n2}). Note that $\mathsf{Y}_{\nu }^{\mu }$ is diffeomorphism invariant,
i.e. under transformation (\ref{dg}) we have $\delta \mathsf{Y}=\mathcal{L}%
_{\varepsilon }\mathsf{Y}$, because of the same $\varepsilon ^{\rho }$ for
all metrics. To calculate invariants of the tensor (\ref{yg}) we take powers
of traces of the the matrix $Y$ corresponding to the tensor $\mathsf{Y}_{\nu
}^{\mu }$, and the number of invariants in 4 dimensions is 4 by the Cayley
theorem, which can be taken as%
\begin{equation}
\varkappa _{1}=\mathrm{Tr}~Y,\ \ \ \varkappa _{2}=\mathrm{Tr}~Y^{2},\ \ \
\varkappa _{3}=\mathrm{Tr}~Y^{3},\ \ \ \varkappa _{4}=\mathrm{Tr}~Y^{4}.
\end{equation}%
Let $\lambda _{\left( i\right) }$ ($i=0,1,2,3$) are eigenvalues of the
tensor $\mathsf{Y}_{\nu }^{\mu }$, which can be treated as relative
eigenvalues of the metric $\mathsf{g}^{\left( 1\right) }$ relatively $%
\mathsf{g}^{\left( 2\right) }$. In the special bi-orthogonal vierbein $%
e_{\mu }^{\left( i\right) }$ the metrics can be written as follows%
\begin{eqnarray}
\mathsf{g}_{\mu \nu }^{\left( 1\right) } &=&\lambda _{\left( 0\right)
}e_{\mu }^{\left( 0\right) }e_{\nu }^{\left( 0\right) }-\lambda _{\left(
1\right) }e_{\mu }^{\left( 1\right) }e_{\nu }^{\left( 1\right) }-\lambda
_{\left( 2\right) }e_{\mu }^{\left( 2\right) }e_{\nu }^{\left( 2\right)
}-\lambda _{\left( 3\right) }e_{\mu }^{\left( 3\right) }e_{\nu }^{\left(
3\right) }, \\
\mathsf{g}_{\mu \nu }^{\left( 2\right) } &=&e_{\mu }^{\left( 0\right)
}e_{\nu }^{\left( 0\right) }-e_{\mu }^{\left( 1\right) }e_{\nu }^{\left(
1\right) }-e_{\mu }^{\left( 2\right) }e_{\nu }^{\left( 2\right) }-e_{\mu
}^{\left( 3\right) }e_{\nu }^{\left( 3\right) },
\end{eqnarray}%
and so the matrix $Y$ is diagonal. In case of real and positive eigenvalues $%
\lambda _{\left( i\right) }$ it is convenient to introduce%
\begin{equation}
\mu _{\left( i\right) }=\mathrm{ln}~\lambda _{\left( i\right) }
\end{equation}%
and consider their powers%
\begin{eqnarray*}
\sigma _{1} &=&\mu _{\left( 0\right) }+\mu _{\left( 1\right) }+\mu _{\left(
2\right) }+\mu _{\left( 3\right) },\ \ \ \ \sigma _{2}=\mu _{\left( 0\right)
}^{2}+\mu _{\left( 1\right) }^{2}+\mu _{\left( 2\right) }^{2}+\mu _{\left(
3\right) }^{2}, \\
\sigma _{3} &=&\mu _{\left( 0\right) }^{3}+\mu _{\left( 1\right) }^{3}+\mu
_{\left( 2\right) }^{3}+\mu _{\left( 3\right) }^{3},\ \ \ \ \sigma _{4}=\mu
_{\left( 0\right) }^{4}+\mu _{\left( 1\right) }^{4}+\mu _{\left( 2\right)
}^{4}+\mu _{\left( 3\right) }^{4}.
\end{eqnarray*}%
Then the scalar potential of bigravity is a function of the introduced
invariants $\sigma _{n}$ as
\begin{equation}
V(\mathsf{g}^{\left( 1\right) },\mathsf{g}^{\left( 2\right) })=\hat{V}%
(\sigma _{1},\sigma _{2},\sigma _{3},\sigma _{4}).
\end{equation}%
An important class of bigravity models has the symmetry $\mathsf{g}^{\left(
1\right) }\leftrightarrow \mathsf{g}^{\left( 2\right) }$. In this case for
eigenvalues we have $\lambda _{\left( i\right) }\rightarrow \lambda _{\left(
i\right) }^{-1}$ and $\mu _{\left( i\right) }\rightarrow -\mu _{\left(
i\right) }$, therefore%
\begin{equation}
\hat{V}(\sigma _{1},\sigma _{2},\sigma _{3},\sigma _{4})=\hat{V}(-\sigma
_{1},\sigma _{2},-\sigma _{3},\sigma _{4}).
\end{equation}

In the weak field limit $\sigma _{i}\rightarrow 0$ it is sufficient to take
into account only two first invariants $\sigma _{1}$, $\sigma _{2}$ and
consider $\hat{V}_{0}\left( \sigma _{1},\sigma _{2}\right) =\hat{V}(\sigma
_{1},\sigma _{2},0,0)$, which appears naturally in brane models \cite%
{kog/mou/pap} and \textquotedblleft Pauli-Fierz-like\textquotedblright\
bigravity \cite{dam/kog/pap}. For the latter we expand%
\begin{equation}
\mathsf{g}_{\mu \nu }^{\left( 1\right) }=\eta _{\mu \nu }+\sqrt{2k_{1}}%
\mathsf{h}_{\mu \nu }^{\left( 1\right) },\ \ \ \mathsf{g}_{\mu \nu }^{\left(
2\right) }=\eta _{\mu \nu }+\sqrt{2k_{2}}\mathsf{h}_{\mu \nu }^{\left(
2\right) },
\end{equation}%
where $\eta _{\mu \nu }$ is the same flat metric. In this limit the mixed
tensor (\ref{yg}) is%
\begin{equation}
\mathsf{Y}_{\nu }^{\mu }=\delta _{\nu }^{\mu }+\sqrt{2k_{1}}\mathsf{h}_{\nu
}^{\left( 1\right) \mu }-\sqrt{2k_{2}}\mathsf{h}_{\nu }^{\left( 2\right) \mu
}.
\end{equation}%
Let us consider the combinations%
\begin{equation}
\mathsf{h}_{\mu \nu }^{0}=q_{1}\mathsf{h}_{\mu \nu }^{\left( 2\right) }+q_{2}%
\mathsf{h}_{\mu \nu }^{\left( 1\right) },\ \ \ \ \ \ \mathsf{h}_{\mu \nu
}^{mass}=q_{1}\mathsf{h}_{\mu \nu }^{\left( 2\right) }-q_{2}\mathsf{h}_{\mu
\nu }^{\left( 1\right) },
\end{equation}%
where $q_{1}^{2}+q_{2}^{2}=1$, then it can be shown that $\mathsf{h}_{\mu
\nu }^{0}$ is massless and $\mathsf{h}_{\mu \nu }^{mass}$ contains the
Pauli-Fierz term. Indeed,%
\begin{eqnarray}
\sigma _{1} &=&\sqrt{2k_{1}}\mathsf{h}_{\mu }^{\left( 1\right) \mu }-\sqrt{%
2k_{2}}\mathsf{h}_{\mu }^{\left( 2\right) \mu }+k_{2}\mathsf{h}_{\mu \nu
}^{\left( 2\right) }\mathsf{h}^{\left( 2\right) \mu \nu }-k_{1}\mathsf{h}%
_{\mu \nu }^{\left( 1\right) }\mathsf{h}^{\left( 1\right) \mu \nu }, \\
\sigma _{2} &=&2k_{1}\mathsf{h}_{\mu \nu }^{\left( 1\right) }\mathsf{h}%
^{\left( 1\right) \mu \nu }+2k_{2}\mathsf{h}_{\mu \nu }^{\left( 2\right) }%
\mathsf{h}^{\left( 2\right) \mu \nu }-4\sqrt{k_{1}k_{2}}\mathsf{h}_{\mu \nu
}^{\left( 1\right) }\mathsf{h}^{\left( 2\right) \mu \nu }.
\end{eqnarray}%
Finally we obtain%
\begin{equation}
\mathsf{h}_{\mu \nu }^{mass}\mathsf{h}^{mass,\mu \nu }-\left( \mathsf{h}%
_{\mu }^{mass,\mu }\right) ^{2}=\dfrac{1}{2\left( k_{1}+k_{2}\right) }\left(
\sigma _{2}-\sigma _{1}^{2}\right) .
\end{equation}%
Thus, if we choose the interaction in the form%
\begin{equation}
S_{int}=-\dfrac{1}{k_{1}+k_{2}}\int d\Omega _{int}\hat{V}_{0}\left( \sigma
_{1},\sigma _{2}\right) ,
\end{equation}%
where $d\Omega _{int}$ is defined in (\ref{n2}) and the scalar interaction
potential is%
\begin{equation}
\hat{V}_{0}^{PF}\left( \sigma _{1},\sigma _{2}\right) =\dfrac{m_{PF}^{2}}{8}%
\left( \sigma _{2}-\sigma _{1}^{2}\right) ,  \label{v0}
\end{equation}%
then the weak field limit of bigravity generates the Pauli-Fierz mass term
\cite{fie/pau} of the shape%
\begin{equation}
S_{int}=-\dfrac{m_{PF}^{2}}{4}\int d^{4}x\left( \mathsf{h}_{\mu \nu }^{mass}%
\mathsf{h}^{mass,\mu \nu }-\left( \mathsf{h}_{\mu }^{mass,\mu }\right)
^{2}\right) .
\end{equation}%
For the brane motivated bigravity scenario \cite{kog/ros,kog/mou/pap/ros/san}
the scalar potential has the form \cite{dam/kog}%
\begin{equation}
\hat{V}_{0}^{brane}\left( \sigma _{1},\sigma _{2}\right) =m^{2}\left(
\mathrm{cosh}~\dfrac{\sigma _{1}}{4}-\mathrm{cosh}~\dfrac{\sqrt{4\sigma
_{2}-\sigma _{1}^{2}}}{4\sqrt{3}}\right) .
\end{equation}%
In the weak field limit it reproduces the Pauli-Fierz mass term (\ref{v0}),
indeed%
\begin{equation}
\hat{V}_{0}^{brane}\left( \sigma _{1},\sigma _{2}\right) |_{m=\sqrt{3}%
m_{PF}}=\hat{V}_{0}^{PF}\left( \sigma _{1},\sigma _{2}\right) .
\end{equation}

Note that the \textquotedblleft perturbative limit\textquotedblright\ which
corresponds to existence of critical point of potential and from which for
bigravity (with potential of form (\ref{yg}) only) it follows that $\mathsf{g%
}_{\mu \nu }^{\left( 1\right) }=\mathsf{g}_{\mu \nu }^{\left( 2\right) }$,
was considered in \cite{dam/kog}. Here we present a more general case which
is not connected with any concrete form of the interaction potential and
does not demand consideration of spaces with constant curvature.

\section{Conclusions}

So in this paper we have analyzed the generalized structure of the
interaction term of multigravity. We introduced the coincidence limit and
obtained the compatibility equation for the interaction potential which was
studied in the weak perturbation limit. We considered the most general
properties of invariant volume and the scalar potential. As an example, we
derived the Pauli-Fierz mass term for bigravity in the weak field limit.

It would be interesting to consider the introduced coincidence limit
in connection with symmetries of the theory and solve the compatibility equation
(\ref{comp})
for concrete models.

\medskip

One of the authors (S.D.) would like to thank V. P. Akulov, J. Bagger, Yu. L.
Bolotin, S. F. Prokushkin, M. D. Schwartz, V. A. Soroka, Yu. P. Stepanovsky and A. V. Vilenkin for
useful and stimulating discussions.

\end{document}